\documentclass[twoside]{article}
\usepackage{actes}
\usepackage{times}
\usepackage{amsmath,amssymb}

\title{Length-Based Attacks for Certain Group Based Encryption Rewriting Systems}
\author{J. Hughes$^1$ \& A. Tannenbaum$^2$}

\titlehead{Length-Based Attacks for Certain Group Based Encryption Rewriting Systems} 
\authorhead{Hughes \& Tannenbaum}     

\affiliation{\begin{tabular}{rr}        
    \\ 1:  Storage Technology Corporation,
    \\     7600 Boone Ave No
    \\     Minneapolis, MN 55428 USA
    \\     {\tt jim@network.com} 
    \\ 2:  Department of Electrical and Computer Engineering,
    \\     Georgia Institute of Technology, 
    \\     Atlanta, GA 30332-0250 USA
    \\     {\tt tannenba@ece.gatech.edu}
  \end{tabular}}

\begin{document}
\setcounter{page}{1}
\maketitle

\begin{abstract} 

In this note, we describe a probabilistic attack on
public key cryptosystems based on the word/conjugacy problems for
finitely presented groups of the type proposed
recently by Anshel, Anshel and Goldfeld.
In such a scheme, one makes use of the property that in the given
group the word problem has a polynomial time solution, while the
conjugacy problem has no known polynomial solution. An example is the
braid group from topology in which the word problem is solvable in
polynomial time while the only known solutions to the conjugacy
problem are exponential.  The attack in this paper is based on having
a canonical representative of each string relative to which a length
function may be computed. Hence the term {\em length attack.} Such
canonical representatives are known to exist for the braid group.

\end{abstract}


\section{Introduction}

Recently, a novel approach to public key encryption based on the
algorithmic difficulty of solving the word and conjugacy problems for
finitely presented groups has been proposed in
\cite{Anshel,AnshelRSA,Ko_crypto,KoCrypto}. The method is based on
having a canonical minimal length form for words in a given finitely
presented group, which can be computed rather rapidly, and in which
there is no corresponding fast solution for the conjugacy problem.  A
key example is the braid group. In this note, we will indicate a
possible probabilistic attack on such a system, using the length
function on the set of conjugates defining the public key.  Note that
since each word has a canonical representative, the length function is
well-defined and for the braid group can be computed in polynomial
time in the word length according to the results in
\cite{Birman-Ko-Lee}.  The attack may be relevant to more general
types of string rewriting cryptosystems, and so we give some of the
relevant background.  Thus this note will also have a tutorial
flavor.

The contents of this paper are as follows.  In Section 2, we make some
general remarks are rewriting systems, and the notion of "length" of a
word.  In Section 3, we define the Artin and Coxeter groups.  In
Section 4, we discuss the classical word and conjugacy problems for
finitely presented groups. In Section 5, the braid cryptosystem of
\cite{Anshel} is described. In Section 6, we give the length attack
for possibly compromising such a cryptosystem, and finally in Section
7 we draw some general conclusions, and directions for further
research for group rewriting based encryption systems.

\section{Background on Monoid and Group Based Rewriting Systems}

In this section, we review some of the relevant concepts from group
theory for rewriting based encryption. We work in this section over a
monoid, but similar remarks hold for group based rewriting systems as
well. 

Let $k$ be an arbitrary field, and $S = \{a_1, \ldots, a_n \}$ a finite
set. Let $S^*$ be the finite monoid generated by $S$, that is, $$S^* =
\{{a_{\sigma(1)} ^{i_1}} \cdots {a_{\sigma(n)} ^{i_n}} \}.$$ Elements
of $S^*$ are called {\em words.}
 We then define the {\em free
algebra} generated by $S$ to be $$A= k[S^*] = k<S>= 
\{\sum_{\sigma \in \Sigma_n} k_{{i_1}
\ldots {i_n}} a_{\sigma(1)} ^{i_1} \cdots a_{\sigma(n)} ^{i_n} \},$$
where $\Sigma_n$ denotes the symmetric group on $n$ letters.

We are now ready to define precisely the key notion of {\em rewriting
system}.  Let $R \subset S^* \times S^*$. We call $R$ the set of
replacement rules.  Many times the pair $(u,v) \in R$ is denoted by $u
\rightarrow v.$ The idea is that when the word $u$ appears inside a
larger word, we replace it with $v$.  More precisely, for any $x,y \in
S^*$, we write $$xuy \rightarrow xvy,$$ and say that the word $xuy$
has been {\em re-written} or {\em reduced} to $xuy.$ $x$ is {\em
irreducible} or {\em normal} if it cannot be rewritten.

We will still need a few more concepts. We say that the {\em rewriting
system} $(S,T)$ is {\em terminating} if there is no infinite chain $x
\rightarrow x_1 \rightarrow x_2 \rightarrow \cdots $ of re-writings.
We then say that the partial ordering $x \ge y$ defined by $x
\rightarrow \cdots \rightarrow y$ is {\em well-founded}.  $R$ is {\em
confluent} if a word $x$ which can be re-written in two different ways
$y_1$ and $y_2$, the re-writings $y_1$ and $y_2$ can be re-written to
a common word $z$.

Note that if $R$ is terminating, confluence means that there exists a
unique irreducible word, $x_{red}$ representing each element of the
monoid presented by the re-writing system. Such a system is called
{\em complete.} Given a word $x \in S^*$, we define the {\em length
of} $x$ or $\ell(x)$, to be the number of generators in $x_{red}$. 

\medskip \noindent {\bf Remark:}\\ In the case of groups, the basic
outline just given is valid.  A key example of a group in which one
can assign a length function is the braid group via the results in
\cite{Birman-Ko-Lee}.  This is the basis of the cryptosystem proposed
in \cite{Anshel}.

\section{Artin and Coxeter Groups}

In this section, we review some of the pertinent background on Artin
and Coxeter groups. An excellence reference for this material in
\cite{Birman}, especially for the braid groups.

Let $G$ be a group. For $a,b \in G$ we define $$<ab>^q := aba \ldots ,
\quad \mbox{product with $q$ factors.}$$ For example, $$<ab>^3 := aba,
\; <ab>^4 :=abab, \; <ba>^5 := babab.$$ An {\em Artin group} is a
group $G$ which admits a set of generators $\{a_i\}_{i\in I}$ with $I$
a totally ordered index set, and with relations of the form $$<a_i a_j
> ^{m_{ij}} = <a_j a_i>^{m_{ji}},$$ for any $i,j \in I$ and with
$m_{ij}$ non-negative integers.  The matrix $M := [m_{ij}]_{i,j \in
I}$ is called the {\em Coxeter matrix.} 

The {\em braid group}, $B_n$, is defined by taking the indexing set $I
:= \{ 1, \ldots, n \}$, and \begin{eqnarray*} m_{ij} &=& 2 \quad
\mbox{for $|i-j| >1$}, \\ m_{i, i+1} &=& m_{i+1, i} =3.
\end{eqnarray*} Thus the  {\em braid group} $B_n$ is a special case of
an Artin group defined by the generators $\sigma_1, \ldots, \sigma_n$,
with the relations \begin{eqnarray*} \sigma_i \sigma_j &=& \sigma_j
\sigma_i \quad |i-j|>1, \; i, j \in I,\\ \sigma_i \sigma_{i+1}
\sigma_i &=& \sigma_{i+1} \sigma_i \sigma_{i+1}.  \end{eqnarray*}

Given an Artin group $G$ with Coxeter matrix $M := [m_{ij}]_{i,j \in
I}$ the associated {\em Coxeter group} is defined by adding the
relations $a_i^2 =1,$ for $i \in I.$ One can easily show them that a
Coxeter group is equivalently defined by the relations $$ (a_i a_j
)^{m_{ij}} =1, \quad i,j \in I, \; \mbox{with} \, m_{ii} =2.$$ 

Artin groups and their associated Coxeter groups have some nice
properties which could make them quite useful in potential rewriting
based systems as we will now see.

\section{Word and Conjugacy Problems for Finitely Presented Groups}

Let $$G = < s_1, s_2, \cdots, s_n : r_1, \cdots, r_k>$$ be a finitely
presented group. Let $U$ be the free monoid generated by $s_i$ and
$s_i^{-1}$.  Then the {\em word problem} is given two strings (words),
$u, v \in U$, decide if $u=v$ in $G$. The {\em conjugacy problem} is
to decide if there exists $a \in G$ such that $u= ava^{-1}$, i.e., $u$
and $v$ are conjugates.

It is well-known that both these problems are algorithmically
unsolvable for general finitely presented groups.  However, for some
very important groups they are solvable, e.g., for Artin groups with
finite Coxeter groups. In fact, Brieskorn and Saito \cite{Brieskorn}
give an explicit solution to the word and conjugacy problems for this
class of groups.  Their algorithm runs in exponential time however.
See also \cite{Eriksson,Eriksson1} and the references therein for some
recent results on the word and conjugacy problems for Coxeter groups.

In some recent work, Birman-Ko-Lee \cite{Birman-Ko-Lee} show that for
the braid group, the word problem is solvable in polynomial time (in
fact, it is quadratic in the word length). Given the results just
described, it has been conjectured that the techniques of
\cite{Birman-Ko-Lee} are extendable to Artin groups with finite
Coxeter groups.  For another solution to this problem see
\cite{Dehornoy}.

At this point, there is no known polynomial time algorithm known for
the conjugacy problem, as originally posed by Artin \cite{Artin}, for
the braid group with $n \ge 6$ strands; see \cite{Birman-Ko-Lee}. It
seems that it is the 
possible complexity of this form of the conjugacy problem which is the
basis of the 
claim of security made
by the authors of the braid cryptosystem in \cite{Anshel}.
(The original conjugacy problem posed by Artin is a decision problem.
Given $x,y \in {B}_n$, is there an $a$ such that $x=a^{-1}xa$?
In the proposed cryptosystems, the public and private keys are known
to be conjugates, so these systems are not based on such a decision
problem. )


For the braid group itself, little work has been accomplished on the
lower and average bounds of the conjugacy search problem for known
conjugates (as in \cite{KoCrypto}) or a system of known conjugates (as
in \cite{AnshelRSA}). There are no proofs that the conjugacy problem
is hard all the time. The motivation to do any of this work has only
occurred recently because these cryptosystems have been proposed.
Some of this work includes a brief look at the probabilities of
colored Burau representation \cite{Hahn}, and other work attempting to
demonstrate the average complexity of the conjugacy problem
\cite{Shpilrain, Shpilrain1} using a set measurement techniques for infinite
groups \cite{Borovik}. Other work has begun on
calculating the normalizer set to solve the conjugacy problem
\cite{Franco} (but this does not help solve the crypto-problem because it
assumes a known conjugator exists.) 

It is important to note 
that there are some important linear representations of the braid group 
namely, the Burau, the colored Burau and the Lawrence-Krammer.
In \cite{AnshelRSA}, it is suggested that the colored Burau 
representation made be used to quickly
solve the word problem.
The Burau representation was originally formulated to prove that the braid
group was linear, but it now is known to have a non-trivial kernel
and as such, cannot be used to solve the general conjugacy problem.
Finally, using a more general representation
due to Lawrence-Krammer, it has been proven that the braid group is
indeed linear \cite{Bigelow}.
This allows linear algebraic methods to be used now in
studying the word and conjugacy problems, and possibly could lead to
yet another attack on braid cryptosystems.  

Finally, note that if one can find a unique irreducible word
from which one can derive a length function, then one can give a
natural distance between words in a given group $G$.  Indeed, let
$\alpha, \beta, \gamma$ denote words relative to a finite presentation
of the group $G$. Let $\ell$ denote the length function which we
assume exists. Then we define the {\em distance} $d$ between the words
$\alpha, \beta$ as $$d_G( \alpha, \beta) := \ell(\alpha \beta^{-1}).$$
It is trivial to check that $d_G$ is a distance function function
between words. See also \cite{Eriksson}.  We will see that this is the
case for the braid group via the results of Birman-Ko-Lee
\cite{Birman-Ko-Lee}.

\section{Braid Cryptosystem}

In some very interesting recent work, Anshel et al.
\cite{Anshel,AnshelRSA}  propose a new twist to rewriting systems for
public key encryption.  We will first state their approach over a
general group.  We should first note however that the use of the word
and conjugacy problems for public-key cryptosystems is not new. An
early reference is \cite{Wagner}.

The general idea is as follows: Alice ($A$) and Bob ($B$) have as
their public keys subgroups of a given group $G$, $$S_A = <s_1,
\ldots, s_n>, \quad <t_1, \ldots t_m>.$$ $A$ chooses a secret element
$a \in S_A$ and $B$ chooses a secret element $b \in S_B$.  $A$
transmits the set of elements $a^{-1} t_1 a, \ldots, a^{-1} t_m a$ and
$B$ transmits the set of elements $b^{-1} s_1 b , \ldots, b^{-1} s_n
b$.  (The elements are rewritten is some fashion before
transmission.)

Now suppose that $$a= {s_{\sigma(1)} ^{i_1}} \cdots {s_{\sigma(n)}
^{i_n}}.$$ Then note that 
\begin{eqnarray*} b^{-1} a b &= & b^{-1} {s_{\sigma(1)} ^{i_1}} \cdots
{s_{\sigma(n)} ^{i_n}}  b \\ &=&  b^{-1} {s_{\sigma(1)} ^{i_1}} b \
b^{-1} {s_{\sigma(2)} ^{i_2}} b \cdots b^{-1} {s_{\sigma(n)} ^{i_n}}b
\\ &= & (b^{-1} {s_{\sigma(1)}} b)^{i_1} \cdots  (b^{-1}
{s_{\sigma(n)}} b)^{i_n} .  \end{eqnarray*} (The conjugate of the
product of two elements is the product of the conjugates.) Thus  $A$
can compute $b^{-1} a b$, and similarly $B$ can compute $a^{-1} b a$.
The common key then is $$a^{-1} b^{-1} ab = [a,b],$$ the commutator of
the two secret elements.

Note that since the two users have the common key written in different
forms, in order to extract the message, it must be reduced to an
identical group element. For the braid group, this can be accomplished
by reducing the commutator to the Birman-Ko-Lee canonical form
\cite{Birman-Ko-Lee}, colored Burau \cite{AnshelRSA} or Dehornoy
\cite{Anshel}.

The braid group is particularly attractive for this protocol since one
has a quadratic time solution for the word problem, and the only known
solution to the conjugacy problem is exponential.

\bigskip \noindent {\bf Remark:}

\medskip \noindent The key properties that underlie this cryptosystem
are having a group in which the word problem is easy to solve (and in
fact each word has a canonical form) and in which the conjugacy
problem is difficult (at least via known techniques). The canonical
form is important as well since it allows a simple method for the
extraction of the common key.

\subsection{Another Braid Cryptosystem}

Another possible cryptosystem based on the word and
conjugacy problems in the braid group has been proposed in
\cite{Ko_crypto,KoCrypto}. In this case, the authors propose the following
scheme: Consider the braid group $B_{n+m}$ on $n+m$ braids.  One
considers two subgroups: $LB_n$ generated by $\sigma_1, \ldots,
\sigma_{n-1}$ and $RB_m$ generated by $\sigma_{n+1}, \ldots
\sigma_{n+m-1}.$ Note that given $a \in LB_n$ and $b \in RB_m$,
$ab=ba$. This is essential for their scheme. 

The protocol for creating a common key then works as follows: The
public key is a pair of integers $(n,m)$, and braid $x \in B_{n+m}$.
Alice choose a secret element $a \in  LB_n$ and sends $axa^{-1}$ to
Bob. Bob chooses a secret element $b \in RB_m$, and sends $bxb^{-1}$
to Alice. Alice can compute $a (bx b^{-1}) a^{-1}$ and Bob can compute
$b (axa^{-1} )b^{-1}$.  Since $a$ and $b$ commute this is the common
key. Being able to solve the Generalized Conjugacy Problem would be
enough to break this system.  It is not known if the converse is
true.

\bigskip \noindent {\bf Remark:}\\ It is an interesting open question
to see if the length attack proposed below may be suitably modified to
be relevant to the protocol in \cite{Ko_crypto}.  It may be also be of
interest to consider some the strong convergent game-theoretic
techniques in \cite{Eriksson,Eriksson1} to study this protocol as well
as that in \cite{Anshel}.

\section{The Length Attack}

In this section, we describe the length attack on word/conjugacy based
encryption systems of the type proposed in \cite{Anshel} in which one
can associate a canonical representative, and therefore a length
function $\ell$ of the type described above. For concreteness, we focus
on the braid group here which has a canonical length function as noted
above.
{\em We should note that the arguments of this section are speculative,
and certainly not mathematically rigorous.}

Research on the length of random words has been
done in the mathematical physics community where braid group has been
valuable in studying certain physical phenomena
\cite{Comtet,Desboisy,Vershik}.  Recall that a symmetric
random walk on a free group $\Gamma_n$ with $n$ generators is
a cross product of a nonsymmetric N-step random walk on a
half--line ${\mathbb{Z}}^+$ and a layer over $N \in {\mathbb{Z}}^+$ 
giving a set of all words
of length $N$ with the uniform distribution (see \cite{Vershik} for the
details). The transition probabilities
in a base are:

$$
N \rightarrow \left\{ 
\begin{array}{l} 
    N + 1 \; \mbox{with the probability} ø\;  \frac{2n-1}{2n} \\ 
    \\
    N - 1 \; \mbox{with the probability} \;  \frac{1}{2n}   
\end{array}
\right.
$$
It is easy to show then that the expectation of a word's length
after $N$ steps is $$N\frac{n-1}{n}$$ and hence the drift is
$$\frac{n-1}{n}.$$ In \cite{Vershik},
the authors show that while the statistical properties of random walks
(Markov chains) on locally free and braid groups are not the same
as uniform statistics on these groups, nevertheless the statistical
characteristics stabilize as the number of generators $n$ grows.
>From this fact, for large $n$ (see \cite{Vershik}, Theorem 11)
given two generic words $x,y \in {B}_n$, the length of
$xy$  will be approximately $\ell(x)+ \ell(y)$. 
(The genericity is important here. For example, if
$x=y^{-1}$ then
$\ell(xy)=0$.) This does give some statistical backing to the length
attack which we are about to formulate.

Given $x \in B_n,$ we say that $y$ is a {\em reducing with respect to 
$x$} (or a {\em reducing element} if $x$ is understood),
if 
$$
\ell(y^{-1}xy) < \ell(x).
$$
The remainder of this discussion will be a way of using substantial
reducing strings in a {\em length attack}, and calculating an upper bound for
the actual difficulty of this attack.  It is important to emphasize
that the ability of removing large reducing elements is not a general solution
to the braid conjugacy problem. It is a specific attack on
word/conjugates encryption systems of the type defined \cite{Anshel}.
Indeed, for such cryptosystems one has the some key information about
the secret elements, namely, the {\em factors are known and their
number bounded.}

Let $$a \in S_A = <s_1, \ldots, s_n>,$$ be the secret element.  Recall
that in the above protocol, $a^{-1} t_r a$ and $t_r$ ($r= 1, \ldots
m$) are publicly given.  We also assume that the factors $s_i$ have
lengths large relative to $a$. For given $r$, set $$u_r = a^{-1} t_r
a.$$ Then the idea is to compute $$\ell (s_j^{\pm 1} u_r s_j^{\mp
1}),$$ repeatedly.  If $s_j$ is a reducing string with respect to $u_r,$ then
one has found a correct factor of $a$ with a certain probability which
will depend on the lengths $\ell(s_i)$ for $i=1, \ldots, n.$ The key is
that the canonical lengths  $\ell(s_i)$ should be large.  
In this case, there is
the greatest probability of a reducing string being formed which can be used to
glean information about $a$.

We can estimate the workload in carrying out such a procedure.
Without loss of generality we can assume that $a$ is made up of $n$
distinct factors combined in $d$ ways. If the length of the $s_i$ is
large, then one join a small number of these factors together to
create a substantial reducing string.  If we include the inverses of the
generators, we should consider $2n$ factors.  Let us call the number
of factors necessary to make a reducing string $k.$ Thus we can create
$(2n)^k$ reducing factors to try. 

By trying all reducing elements, a pattern that there are certain factors which
annihilate better than others should be observed.  One can do this on
a single public conjugate in $(2n)^k$ operations.  This pattern can be
significantly reinforced by repeating this $n$ times on each public
conjugate $a t_r a^{-1}.$ Combining all the steps above brings us to
$n{(2n)}^k$ operations.

Relative to the lengths $\ell(s_i)$ of the generators $s_i$ (and the
specific group chosen), we conjecture that in a number of cases this
will be sufficiently reliable to removing a given $s_i,$ so that
backtracking will not be necessary. We can now do this $dn$ times
bringing the total to $dn(2n)^k$ operations. 

This is polynomial to the number of different factors, and linear to
the number of factors in the public keys.  This is the basis of the
length attack.

Another demonstration of this idea is trivial. If one sets $k=d$ then
the first of the $dn$ passes will solve the system in the expected
exponential time $2n^k$ steps. This is simply an enumeration of all
possible values of $a$. If one sets $k=1$ then, if individual factors
are not significant, this attack will not work. If there is a value of
$k<d$ that works, this attack significantly reduces the strength of
the result. Once this attack is valid, then lengthening the private
key only linearly increases the time to solution.

Depending on the values chosen for the cryptosystem in \cite{Anshel},
$k$ may need to be longer than the actual word $a$, as has been
suggested in \cite{AnshelRSA}\footnote{This attack was known to the
authors before that paper was written}. Yet another potential
problem is that if the
factors are simple, other attacks such as those proposed in
\cite{HughesLinear} may be effective.

In some sense, the length attack is reminiscent of the ``smoothness''
attack for the Diffie-Hellman public key exchange system based on the
discrete logarithm \cite{Schneier}. In this case, the protocol may be
vulnerable when all of the prime factors of $q-1$ (where the base
field for the discrete logarithm has $q$ elements) are small. (Such a
number is called {\em smooth}.)

\section{Conclusions}

We have made a computation which indicates that a length attack on a
conjugacy/word problem cryptosystem of the type defined in
\cite{Anshel} has difficulty bounded above by $dn{(2n)}^k.$ Given this
conjecture, the only exponential aspect is the number of factors
necessary to form a reliable reducing string. To make this secure, $k$ needs to
be 100 or larger. 

In addition, as described, this attack does not use many tricks that
one can use in order to speed up this length algorithm by several
orders of magnitude.  These include randomized and/or genetic
algorithms which lead to more probabilistic solutions. 

The bottom line is that the length attack forces one to take
generators of not too long canonical length.
Dorian Goldfeld reports that experimental evidence suggests that if each
of the generators $s_1, \ldots, s_n$ is of length $<10$ in the Artin
generators, then this may foil the attack. All of this still must be
tested.

Finally, it is important to note that this attack does not solve the general
conjugacy problem for the braid group. Indeed, in this case the
factors of $a$ are known and bounded. In the general conjugacy
problem, the number of possible factors of $a$ is infinite.
Consequently, the the conjugacy problem seems to be much harder and
not amenable to this technique.  The key exchange of the type proposed
in \cite{Anshel} requires the factors be known and communicated, and
give the attacker far more information than is known to the general
conjugacy problem. 
 
\section{Acknowledgments}

We wish to thank the principals of Arithmetica Inc., Mike and Iris
Anshel and Dorian Goldfeld who introduced some of us to the braid group,
invented this interesting cryptosystem, took the ideas seriously and
whose claims motivate this work.

\pagebreak

\pagebreak 
\thispagestyle{colloquetitle}
\cleardoublepage
\end{document}